# Scattering phase shift for relativistic exponential-type separable potentials

**A. D. Alhaidari**
*Physics Department, King Fahd University of Petroleum & Minerals, Box 5047, Dhahran 31261, Saudi Arabia*
E-mail: **haidari@mailaps.org**

**Abstract:** The J-matrix method of scattering is used to obtain analytic expressions for the phase shift of two classes of relativistic exponential-type separable potentials whose radial component is either of the general form $r^{\nu-1}e^{-\lambda r/2}$ or $r^{2\nu}e^{-\lambda^2 r^2/2}$, where $\lambda$ is a range parameter and $\nu$ = 0, 1, or 2. The rank of these separable potentials is $\nu$ + 1. The nonrelativistic limit is obtained and shown to be identical to the nonrelativistic phase shift. An exact numerical evaluation for higher order potentials ($\nu \geq 3$) can also be obtained in a simple way as illustrated for the case $\nu = 3$.

## I. PRELIMINARIES

Separable potentials are the simplest nontrivial realization of two-point nonlocal potentials that are used to model particle interactions or to implement schemes and methods being introduced on toy models. These potentials form the kernel of an integral operator which couples only very few number of the lowest states of the system. In its simplest form (one-term separable potential) coupling affects only the lowest state and, thus, the matrix representation, $V_{nm}$, of the potential is zero except for $V_{00}$. Normally in such models, exact solutions for the bound and/or scattering states are obtained provided that the $H_0$–problem is analytically soluble, where $H_0$ is the reference Hamiltonian in the absence of the separable potential. Fewer number of exactly soluble models exist for higher rank $M$–term separable potentials, where $M > 2$. Furthermore, even fewer exact solutions exist in the relativistic as opposed to the nonrelativistic regime. In the published literature, separable potentials are used more widely in nuclear and condensed matter than in other fields of physics. The amount of publications dealing with the application of separable potentials to the solution of various problems in physics is remarkable. As an example, we select from the result of a literature survey of the tens of papers dealing with this subject annually articles during the last five years that are of relevance to our present work [1–23].

In this article, we consider relativistic exponential-type separable potentials with spherical symmetry and take Dirac Hamiltonian as the reference Hamiltonian. We investigate two classes of these potentials whose radial component is either of the general form $r^{\nu-1}e^{-\lambda r/2}$ or $r^{2\nu}e^{-\lambda^2 r^2/2}$, where $\lambda$ is a range parameter and $\nu$ = 0, 1, 2, and 3. It turns out that these potentials are ($\nu$ + 1)–term separable, that is their matrix representation is of dimension ($\nu$ + 1). The tools of the relativistic J–matrix method of scattering [24] are used to obtain analytic expressions for the phase shift for $\nu$ = 0, 1, and 2. For higher order potentials ($\nu \geq 3$), we find that it is sufficient and more practical to calculate the <u>exact</u> phase shift numerically using the relativistic J–matrix method with a function space dimension $N \geq \nu + 1$. The resulting structure of the phase shift as a function of energy turns out to be very rich and highly interesting as demonstrated by the graphical results in the given examples.

In atomic units ($m = e = \hbar = 1$) and taking the speed of light $c = \alpha^{-1}$, Dirac equation for a nonlocal potential $V(\vec{r},\vec{r}')$ reads:

$$\begin{pmatrix} 1 & -i\alpha\vec{\sigma}\cdot\vec{\nabla} \\ -i\alpha\vec{\sigma}\cdot\vec{\nabla} & -1 \end{pmatrix}\Phi(\vec{r}) + \alpha^2\int V(\vec{r},\vec{r}')\Phi(\vec{r}')d^3\vec{r}' = \varepsilon\,\Phi(\vec{r}) \tag{1.1}$$

where $\alpha$ is the fine structure constant, $\vec{\sigma}$ are the three 2×2 Pauli spin matrices, and $\varepsilon$ is the relativistic energy. For spherically symmetric potentials, the wave function $\Phi(\vec{r})$ is an element of an $L^2$ space spanned by the four-component spinor basis [25]:

$$\psi_n^{lm}(\vec{r}) = \begin{pmatrix} i\dfrac{\phi_n(r)}{r}\chi_{lm} \\ \dfrac{\theta_n(r)}{r}\vec{\sigma}\cdot\hat{r}\chi_{lm} \end{pmatrix} \quad \begin{matrix} ;n,l = 0,1,2,... \\ ;m = -l\pm½, -l+1\pm½, ..., l\pm½ \end{matrix} \tag{1.2}$$

where $\chi_{lm}$ is the angular spinor component and $\{\phi_n, \theta_n\}$ are the radial spinor components. Therefore, in this basis, Dirac equation (1.1) can be written as

$$\begin{pmatrix} 1-\varepsilon & \alpha\left(\dfrac{\kappa}{r} - \dfrac{d}{dr}\right) \\ \alpha\left(\dfrac{\kappa}{r} + \dfrac{d}{dr}\right) & -1-\varepsilon \end{pmatrix}\begin{pmatrix} \sum_n h_n(\varepsilon)\phi_n(r) \\ \sum_n h_n(\varepsilon)\theta_n(r) \end{pmatrix} + \alpha^2\int V(\vec{r},\vec{r}')\Phi(\vec{r}')d^3\vec{r}' = 0 \tag{1.3}$$

where $\kappa$ is the spin-orbit coupling parameter defined by $\kappa = \pm(j + ½)$ for $l = j \pm ½$, $j$ is the total angular momentum quantum number, and $\{h_n(\varepsilon)\}_{n=0}^{\infty}$ is the set of expansion coefficients of the spinor wave function. For a spherically symmetric relativistic separable potential, the kernel $V(\vec{r},\vec{r}')$, can generally be written as

$$V(\vec{r},\vec{r}') = \begin{pmatrix} V_+U(r)U(r') & V_0U(r)W(r') \\ V_0W(r)U(r') & V_-W(r)W(r') \end{pmatrix} \tag{1.4}$$

where $U(r)$ and $W(r)$ are real radial potential functions; $V_\pm$ and $V_0$ are real constant coupling parameters. It is clear that the representation (1.4) satisfies unitarity, namely $V(\vec{r},\vec{r}')^\dagger = V(\vec{r}',\vec{r})$. The potential matrix elements are

$$V_{nm}^{l,l'} = \alpha^2\iint \psi_n^l(\vec{r})^\dagger V(\vec{r},\vec{r}')\psi_m^{l'}(\vec{r}')d^3\vec{r}d^3\vec{r}' \tag{1.5}$$

This integral is tractable only when $l = 0$ (i.e., $\kappa = 0$). Therefore, our problem is constrained to the S-wave separable potentials. The angular contribution to the integral is a factor of $4\pi$. Hence,

$$V_{nm} = 4\pi\alpha^2\left[V_+I_nI_m + V_-J_nJ_m + V_0\left(I_nJ_m + I_mJ_n\right)\right] \tag{1.6}$$

where we've defined the following integrals

$$I_n = \int rU(r)\phi_n(r)dr \quad \text{and} \quad J_n = \int rW(r)\theta_n(r)dr \tag{1.7}$$

In section II, an analytic expression for the phase shift associated with the S-wave relativistic separable potential (1.4) with $U(r) = W(r) = r^{\nu-1}e^{-\lambda r/2}$ for $\nu = 0,1,2$ is obtained using the J-matrix formalism. While in section III, the same treatment is repeated for the S-wave relativistic separable potential with $U(r) = \lambda rW(r) = r^{2\nu}e^{-\lambda^2r^2/2}$ for $\nu = 0,1,2$. In section IV, we obtain graphical results for the phase shift angle as a function of energy for a given choice of range and spin-dependent coupling parameters. An overview of the essential formalism of the relativistic J-matrix method needed for the present work is given in Appendix A. Readers who are not familiar with the J-matrix method of scattering and interested in more than the overview given in the Appendix can consult [26] and references therein. The relativistic extension of the J-matrix theory can be found in [24,27,28].

## II. THE SEPARABLE POTENTIAL $r^{\nu-1}e^{-\lambda r/2}$

The natural basis for this problem is the relativistic two-component Laguerre functions given by equation (A.7) in Appendix A. For $\kappa$ = 0, the separable potential (1.4) with $U(r)=W(r)=r^{\nu-1}e^{-\lambda r/2}$ will result in a finite number of terms for both integrals in (1.7) simultaneously:

$$\begin{aligned} I_n &= \frac{1}{\lambda^{\nu+1/2}\sqrt{n+1}}\int x^{\nu+1}e^{-x}L_n^1(x)dx \\ J_n &= \frac{\sqrt{n+1}}{2}\frac{\lambda C}{\lambda^{\nu+1/2}}\int x^{\nu}e^{-x}\left[L_n^0(x)+L_{n+1}^0(x)\right]dx \end{aligned} \qquad (2.1)$$

where $x = \lambda r$.

In the case of Yukawa-type separable potential (that is $\nu$ = 0), the first integral in (2.1) reads

$$I_n = \frac{1}{\sqrt{\lambda(n+1)}}\int xe^{-x}L_n^1(x)L_0^1(x)dx \qquad (2.2)$$

since $L_0^{\mu}(x)=1$. Using the normalization property of the Laguerre polynomials [29], we obtain:

$$I_n = \frac{\delta_{n0}}{\sqrt{\lambda}} \qquad (2.3)$$

Similarly, the second integral gives

$$J_n = \frac{C}{2}\sqrt{\lambda}\delta_{n0} \qquad (2.4)$$

Therefore,

$$V_{nm} = 4\pi\frac{\alpha^2}{\lambda}\left[V_+ + \left(\lambda C/2\right)^2 V_- + \lambda C V_0\right]\delta_{n0}\delta_{m0} \qquad (2.5)$$

Hence, the potential matrix is zero except for $V_{00}$. The scattering phase shift associated with this relativistic separable potential can be obtained analytically using the J-matrix method. For our present problem, the reference Hamiltonian is the sum of the kinetic energy term [the free Dirac Hamiltonian given by (A.6)] and this Yukawa-type separable potential $r^{-1}e^{-\lambda r/2}$. This means that we need to find an analytic solution to the newly emerging recursion relation and obtain the sine-like, $h_n = \hat{s}_n$, and cosine-like, $h_n = \hat{c}_n$, expansion coefficients of the regularized wave function necessary for J-matrix calculations. Then we proceed as follows:

The matrix representation of the new reference Hamiltonian $\hat{H}_0$ is the sum of that of the free Dirac Hamiltonian $H_0$, given by (A.9) with $\kappa$ = 0, and the 1×1 matrix representation of the one-term separable potential $V$ in (2.5). Let $\Im_{nm}=\left(H_0\right)_{nm}-\varepsilon\,\Omega_{nm}$, where $\Omega$ is given by (A.10) with $\kappa$ = 0, be the J-matrix for the original Dirac problem without the Yukawa separable potential and let $h_n$ stand for either $\hat{s}_n$ or $\hat{c}_n$ of the current problem. Then, the three-term symmetric recursion relation

$$\Im_{n,n-1}h_{n-1}+\Im_{n,n}h_n+\Im_{n,n+1}h_{n+1}=0\ ;\quad n\geq 1 \tag{2.6}$$

is the same as that of the original problem, which is given by (A.2), except for the initial conditions which now read

$$\begin{aligned}\Im_{00}\hat{s}_0+\Im_{01}\hat{s}_1&=-V_{00}\hat{s}_0\\ \Im_{00}\hat{c}_0+\Im_{01}\hat{c}_1&=-V_{00}\hat{c}_0-\alpha^2 w/2\hat{s}_0\end{aligned} \tag{2.7}$$

where $w(\varepsilon)$ is the Wronskian of the regular and irregular solutions of the free Dirac problem. Now, we seek a linear unitary transformation that mixes the original sine-like and cosine-like solutions, $\{s_n, c_n\}$, as

$$\begin{pmatrix}\hat{s}_n\\ \hat{c}_n\end{pmatrix}=\begin{pmatrix}\gamma & \rho\\ -\rho & \gamma\end{pmatrix}\begin{pmatrix}s_n\\ c_n\end{pmatrix}\qquad ; n\geq 0 \tag{2.8}$$

such that we recover the original initial conditions (A.3) when $V$ = 0. $\gamma$ and $\rho$ are energy dependent and parameterized by $V_{00}$. They are, respectively, the cosine and sine of an angle, say, $\tau$. This angle can be thought of as the phase shift from the original solutions, $s_n(\varepsilon)$ and $c_n(\varepsilon)$, of the unperturbed problem due to this Yukawa-type separable potential. That is, the phase shift of the current problem is obtained, simply, as the rotation angle of the kinematical coefficients of the original relativistic J-matrix problem. We can obtain a more compact and transparent solution if we write the problem in terms of this angle, $\tau$, and the complex coefficients defined by

$$g_n^{\pm}(\varepsilon)=c_n(\varepsilon)\pm i s_n(\varepsilon) \tag{2.9}$$

In this notation, the transformation (2.8) is equivalent to

$$\hat{g}_n^{\pm}=e^{\pm i\tau}g_n^{\pm}\qquad ; n\geq 0 \tag{2.10}$$

The J-matrix Kinematical coefficients $T_n(\varepsilon)$ and $R_n^{\pm}(\varepsilon)$ defined in (A.4) transform, according to (2.10), simply as follows:

$$\hat{T}_n=e^{-2i\tau}T_n\ ;\ \hat{R}_{n+1}^{\pm}=R_{n+1}^{\pm}\qquad ; n\geq 0 \tag{2.11}$$

Moreover, the set of initial conditions in (A.3) are written in terms of the complex coefficients as

$$\Im_{00}g_0^{+}+\Im_{01}g_1^{+}=\frac{-i\alpha^2 w/g_0^{+}}{1-T_0} \tag{2.12}$$

While, the new initial conditions in (2.7) take the following form

$$\Im_{00}\hat{g}_0^{+}+\Im_{01}\hat{g}_1^{+}=-V_{00}\hat{g}_0^{+}\,\frac{-i\alpha^2 w/\hat{g}_0^{+}}{1-\hat{T}_0} \tag{2.13}$$

Substituting from (2.10) into (2.13) and using (2.12) we obtain

$$e^{2i\tau}=T_0+\left(1-T_0\right)\left[1+\frac{V_{00}}{\Im_{00}+\Im_{01}R_1^{+}}\right]^{-1} \tag{2.14}$$

This expression for the phase shift angle $\tau$ can be evaluated using $T_0(\varepsilon)$ and $R_1^{\pm}(\varepsilon)$ given by the set of equations (A.11)-(A.13) in Appendix A. The nonrelativistic limit of the phase shift in (2.14) is obtained by using (A.14). After some manipulations, we obtain:

$$\tan(\tau)=\pm\frac{k}{\lambda}\left\{\left[\frac{1}{4}-\left(\frac{k}{\lambda}\right)^2\right]+\frac{\lambda^3}{8\pi V_+}\left[\frac{1}{4}+\left(\frac{k}{\lambda}\right)^2\right]^2\right\}^{-1} \tag{2.15}$$

This is the well-known nonrelativistic result for this separable potential in the absence of Coulomb interaction [30].

For Yamaguchi-type separable potential (that is $\nu = 1$) the integrals in (2.1) give the following 2×2 symmetric potential matrix after using the recursion and normalization properties of the Laguerre polynomials [29]:

$$\begin{aligned} V_{00} &= 16\pi \frac{\alpha^2}{\lambda^3} V_+ \\ V_{11} &= 8\pi \frac{\alpha^2}{\lambda^3}\left[V_+ + \left(\lambda C/2\right)^2 V_- + \lambda C V_0\right] \\ V_{01} &= V_{10} = -8\sqrt{2}\pi \frac{\alpha^2}{\lambda^3}\left[V_+ + \left(\lambda C/2\right) V_0\right] \end{aligned} \tag{2.16}$$

Again, an analytic solution of this problem can also be obtained in the context of the J-matrix formalism for a reference Hamiltonian which is the sum of the S-wave kinetic energy part, $H_0$ in (A.9) with $\kappa = 0$, and this 2×2 matrix representation of the Yamaguchi-type separable potential $e^{-\lambda r/2}$. Therefore, we proceed as in the previous case and as follows:

The homogeneous recursion (2.6) in this case is valid only for $n \geq 2$, while the initial relations that replace (2.7), or equivalently (2.13), read

$$\begin{pmatrix} \Im_{00} & \Im_{01} & 0 \\ \Im_{10} & \Im_{11} & \Im_{12} \end{pmatrix}\begin{pmatrix} \hat{g}_0^+ \\ \hat{g}_1^+ \\ \hat{g}_2^+ \end{pmatrix} = -\begin{pmatrix} V_{00} & V_{01} \\ V_{10} & V_{11} \end{pmatrix}\begin{pmatrix} \hat{g}_0^+ \\ \hat{g}_1^+ \end{pmatrix} + \frac{-i\alpha^2 w/\hat{g}_0^+}{1-\hat{T}_0}\begin{pmatrix} 1 \\ 0 \end{pmatrix} \tag{2.17}$$

In this case a transformation similar to (2.10) will not be sufficient to recover the original initial condition (2.12), however, the following split transformation will:

$$\begin{aligned} \hat{g}_0^{\pm} &= \eta e^{\pm i\xi} g_0^{\pm} \\ \hat{g}_n^{\pm} &= e^{\pm i\tau} g_n^{\pm} \quad ; n \geq 1 \end{aligned} \tag{2.18}$$

where $\eta$ and $\xi$ are real, and $\eta > 0$. Again the angle $\tau$ is the phase shift from the original sine-like and cosine-like solutions of the unperturbed problem due to this Yamaguchi-type separable potential. The details of the calculation are left to Appendix B with the following end result

$$\begin{aligned} e^{2i\tau} = T_0 e^{-2i\zeta} + \left(1-T_0\right)\left(\Im_{01}+V_{01}\right)\left(\frac{\Im_{00}+\Im_{01}R_1^+}{\Im_{01}-V_{11}R_1^+}\right)\times \\ \left[\left(\Im_{01}-V_{11}R_1^+\right)\frac{\Im_{00}+V_{00}}{\Im_{01}+V_{01}} + R_1^+\left(\Im_{01}+V_{01}\right)\right]^{-1} \end{aligned} \tag{2.19}$$

where the angle $\zeta$ is given by equation (B.9) in Appendix B, while $T_0$ and $R_1^{\pm}$ are again given by the equation set (A.11-A.13). The nonrelativistic limit of the phase shift in (2.19) is also obtained by using (A.14).

A similar treatment for the case $\nu = 2$ gives the following expression for the scattering matrix

$$\begin{aligned} e^{2i\tau} = T_0 e^{-2i\xi} + \frac{1-T_0}{R_1^+\Lambda}\left(\Im_{00}+R_1^+\Im_{01}\right)\times \\ \left\{R_1^+\Lambda\left(\Im_{00}+V_{00}\right) + R_1^+\left[\frac{\Im_{01}+V_{01}}{\Im_{12}+V_{12}}(\Im_{12}-R_2^+V_{22}-V_{02}\Lambda)+R_2^+V_{02}\right]\right\}^{-1} \end{aligned} \tag{2.20}$$

where $V_{nm}$ is evaluated using equation (1.6) with

$$I_n = \frac{6}{\lambda^{5/2}}\left(\frac{\delta_{n2}}{\sqrt{3}} - \sqrt{2}\delta_{n1} + \delta_{n0}\right)$$
$$J_n = \frac{\lambda C}{\lambda^{5/2}}\left(\sqrt{3}\delta_{n2} - \sqrt{2}\delta_{n1} - \delta_{n0}\right) \tag{2.21}$$

and

$$\xi = \arg\left(R_1^+ \Lambda\right) \tag{2.22}$$

where

$$\Lambda(\varepsilon, V) = \frac{\left(\Im_{01}/R_1^+\right) + \Im_{11} - R_2^+ V_{12} + \frac{\Im_{11} + V_{11}}{\Im_{12} + V_{12}}\left(-\Im_{12} + R_2^+ V_{22}\right)}{\Im_{01} + V_{01} - V_{02}\,\frac{\Im_{11} + V_{11}}{\Im_{12} + V_{12}}} \tag{2.23}$$

Note that $R_2^+$ can be calculated recursively using the recursion relation as give by (A.5), that is:

$$R_2^{\pm} = -\frac{1}{\Im_{12}}\left(\Im_{11} + \frac{\Im_{01}}{R_1^{\pm}}\right) \tag{2.24}$$

Using the recursion and normalization properties of the Laguerre polynomials [29] one can show that in the general case, the rank of this separable potential is $\nu + 1$. That is the matrix representation of the separable potential in the relativistic Laguerre basis is of dimension $\nu + 1$. For $\nu \geq 3$, the analytic evaluation is too lengthy, however, for numerical J-matrix computations only the matrix representation of the potential is needed. It is our observation that in all cases above the accuracy of the numerical J-matrix calculation, when compared with the analytic results, is limited only by the computing machine. Therefore, it is sufficient and more practical to calculate the <u>exact</u> phase shift for higher order potentials ($\nu \geq 3$) numerically using the relativistic J-matrix method with a function space dimension $N \geq \nu + 1$. In such calculation we only need to evaluate the integrals in equation (2.1). For $\nu = 3$, the evaluation of these integrals give

$$I_n = \frac{-12}{\lambda^{7/2}}\left(\delta_{n3} - 2\sqrt{3}\delta_{n2} + 3\sqrt{2}\delta_{n1} - 2\delta_{n0}\right)$$
$$J_n = -\frac{6\lambda C}{\lambda^{7/2}}\left(\delta_{n3} - \sqrt{3}\delta_{n2} + \delta_{n0}\right) \tag{2.25}$$

Inserting these expressions in equation (1.6) gives the elements of the exact 4×4 potential matrix. These will be added to the $N\times N$ tridiagonal matrix representation of the S-wave reference Hamiltonian [given by (A.9) with $\kappa = 0$] resulting in the $N\times N$ total Hamiltonian. This is then used in the standard relativistic J-matrix calculation of the finite Green's function leading to the phase shift [24,28].

The nonrelativistic phase shift is obtained by the standard nonrelativistic J-matrix method [26] with a short range potential whose matrix elements are $V_{nm} = 4\pi V_+ I_n I_m$, where $I_n$ is as given in (2.25) above.

Higher order potentials can also be handled numerically in the same way.

## III. THE SEPARABLE POTENTIAL $r^{2\nu} e^{-\lambda^2 r^2/2}$

In this section, we repeat briefly the same development that was carried out above for this new Gaussian-type separable potential, however, in a different $L^2$ basis. The natural basis for this problem is the relativistic two-component oscillator-type basis given by equation (A.15) in Appendix A. For $\kappa$ = 0, the separable potential (1.4) with $U(r) = W(r) = r^{2\nu} e^{-\lambda^2 r^2/2}$ will not result in a finite number of terms for both integrals in (1.7) simultaneously. However, the following choice will:

$$U(r) = r^{2\nu} e^{-\lambda^2 r^2/2}$$
$$W(r) = \frac{1}{\lambda r} r^{2\nu} e^{-\lambda^2 r^2/2} \tag{3.1}$$

It gives:

$$I_n = \frac{1/\sqrt{2}}{\lambda^{2\nu+3/2}} \sqrt{\frac{\Gamma(n+1)}{\Gamma(n+3/2)}} \int x^{\nu+1/2} e^{-x} L_n^{1/2}(x) dx$$
$$J_n = \frac{\lambda C/\sqrt{2}}{\lambda^{2\nu+3/2}} \sqrt{\frac{\Gamma(n+1)}{\Gamma(n+3/2)}} \int x^{\nu-1/2} e^{-x} \left[ (n+1/2) L_n^{-1/2}(x) + (n+1) L_{n+1}^{-1/2}(x) \right] dx \tag{3.2}$$

where $x = \lambda^2 r^2$.

In the case of Gaussian potentials (that is $\nu$ = 0), the integrals in (3.2) give the following potential matrix:

$$V_{nm} = \pi\sqrt{\pi} \frac{\alpha^2}{\lambda^3} \left[ V_+ + (\lambda C)^2 V_- + 2\lambda C V_0 \right] \delta_{n0} \delta_{m0} \tag{3.3}$$

To find the analytic expression for the phase shift associated with this separable potential then we proceed as in section II and as follows:

The matrix representation of the new reference Hamiltonian $\hat{H}_0$ is the sum of that of $H_0$, given by (A.17) with $\kappa$ = 0, and the 1×1 potential matrix in (3.3). The same computational steps of section II lead to the same expression (2.14) for the phase shift, however, in terms of $V_{00}$ given by equation (3.3) and the coefficients $T_0$ and $R_1^{\pm}$ which are obtained by substituting $\{s_0, s_1, c_0, c_1\}$ given by (A.19) into (A.4). The nonrelativistic limit of the phase shift is also obtained by using (A.14). This leads to the same formula, (2.14), with the coefficients $T_0$ and $R_1^{\pm}$ being obtained again using (A.19) except that we take $k = \sqrt{2E}$.

For the case $\nu$ = 1, the evaluation of the integral in (3.2) gives the following 2×2 symmetric potential matrix whose elements are:

$$V_{00} = \frac{9}{4} \pi\sqrt{\pi} \frac{\alpha^2}{\lambda^7} \left[ V_+ + (\lambda C/3)^2 V_- - 2(\lambda C/3) V_0 \right]$$
$$V_{11} = \frac{3}{2} \pi\sqrt{\pi} \frac{\alpha^2}{\lambda^7} \left[ V_+ + (\lambda C)^2 V_- + 2(\lambda C) V_0 \right] \tag{3.4}$$
$$V_{01} = V_{10} = -\frac{3\pi}{2} \sqrt{\frac{3\pi}{2}} \frac{\alpha^2}{\lambda^7} \left[ V_+ - \frac{1}{3} (\lambda C)^2 V_- + \frac{2}{3} \lambda C V_0 \right]$$

On the other hand, for $\nu$ = 2, we obtain a 3×3 potential matrix using

$$I_n = \frac{\pi^{1/4}}{2\lambda^{11/2}}\left(\sqrt{\frac{3\times 5}{2}}\delta_{n2} - 5\sqrt{\frac{3}{2}}\delta_{n1} + \frac{3\times 5}{4}\delta_{n0}\right)$$
$$J_n = \frac{\lambda C\pi^{1/4}}{2\lambda^{11/2}}\left(\sqrt{\frac{3\times 5}{2}}\delta_{n2} - \sqrt{\frac{3}{2}}\delta_{n1} - \frac{3\times 3}{4}\delta_{n0}\right) \tag{3.5}$$

The resulting expression for the phase shift angle $\tau$ in these two cases is exactly that given by equation (2.19) for $\nu = 1$ and by equation (2.20) for $\nu = 2$, except that we should use the following parameters instead:

1. The potential matrix elements in (3.4) for $\nu = 1$ or those obtained using (3.5) in (1.6) for $\nu = 2$,
2. The J-matrix elements $\Im_{nm}$ given by those of $H_0$ in (A.17), and those of $\Omega$ in (A.18), both with $\kappa = 0$, and
3. $T_0(\varepsilon)$ and $R_1^+(\varepsilon)$ obtained by substituting the relativistic sine-like and cosine-like coefficients given by (A.19) into (A.4).

The nonrelativistic limit of this phase shift is obtained similarly by using (A.14)

Again, for the general case these potentials are of rank $\nu + 1$. It is sufficient and more practical to calculate the exact phase shift for higher order potentials ($\nu \geq 3$) numerically using the relativistic J-matrix method with a function space whose dimension is $N \geq \nu + 1$. In such calculation we only need to evaluate the integrals in (3.2). For $\nu = 3$ we obtain:

$$I_n = \frac{3\pi^{1/4}}{4\lambda^{15/2}}\left(-\sqrt{7\times 5}\delta_{n3} + 3\sqrt{\frac{3\times 5}{2}}\delta_{n2} - \frac{5\times 7}{2}\sqrt{\frac{3}{2}}\delta_{n1} + \frac{5\times 13}{4}\delta_{n0}\right)$$
$$J_n = \frac{3\lambda C\pi^{1/4}}{4\lambda^{15/2}}\left(-\sqrt{7\times 5}\delta_{n3} + 3\sqrt{\frac{3\times 5}{2}}\delta_{n2} + \frac{5}{2}\sqrt{\frac{3}{2}}\delta_{n1} - \frac{5\times 5}{4}\delta_{n0}\right) \tag{3.6}$$

Inserting these expressions in equation (1.6) gives the exact elements of the 4×4 potential matrix. These will be added to the $N\times N$ tridiagonal matrix representation of the S-wave reference Hamiltonian [given by (A.17) with $\kappa = 0$] resulting in the $N\times N$ total Hamiltonian. This is then used in the standard relativistic J-matrix calculation of the finite Green's function leading to the phase shift.

The nonrelativistic limit is obtained by the standard nonrelativistic J-matrix method with a short range potential whose matrix elements are $V_{nm} = 4\pi V_+ I_n I_m$ with $I_n$ as given in (3.6).

Higher order potentials can also be handled the same way.

## IV. GRAPHICAL RESULTS

The following examples share the same potential parameters:

$$V_+ = 0.5, V_- = 0.3, V_0 = -0.2 \tag{4.1}$$

The separable potential in the first four examples is of the form $r^{\nu-1}e^{-\lambda r/2}$ with $\nu = 0$, 1, 2, and 3, respectively. The last four are concerned with the Gaussian-type $r^{2\nu}e^{-\lambda^2 r^2/2}$ for $\nu = 0$, 1, 2, and 3, respectively. The results are shown graphically as a plot of the phase shift angle, in radians, vs. the nonrelativistic energy, in atomic units. The relativistic phase shift is shown as a solid line while the nonrelativistic one, for the same example, is shown as a dotted line on the same graph. Figure (1) shows the results of the first four examples, while Figure (2) shows those of the last four. The rest of the parameter values are taken as follows:

Figure (1-a): $\nu = 0$, $\lambda = 1.0$, $\alpha = 0.4$, $C = \alpha / 3$
Figure (1-b): $\nu = 1$, $\lambda = 1.5$, $\alpha = 0.5$, $C = \alpha / 3$
Figure (1-c): $\nu = 2$, $\lambda = 2.0$, $\alpha = 0.5$, $C = \alpha / 3$
Figure (1-d): $\nu = 3$, $\lambda = 2.5$, $\alpha = 0.2$, $C = \alpha / 3$

Figure (2-a): $\nu = 0$, $\lambda = 1.0$, $\alpha = 0.5$, $C = \alpha / 3$
Figure (2-b): $\nu = 1$, $\lambda = 1.1$, $\alpha = 0.4$, $C = \alpha / 3$
Figure (2-c): $\nu = 2$, $\lambda = 1.4$, $\alpha = 0.3$, $C = \alpha / 4$
Figure (2-d): $\nu = 3$, $\lambda = 1.5$, $\alpha = 0.3$, $C = \alpha / 4$

Note that in Examples 4 and 8 [respectively, Figures (1-d) and (2-d)] the relativistic as well as nonrelativistic phase shift are calculated numerically using the J-matrix method while the rest are obtained analytically.

## ACKNOWLEDGMENTS

The author is indebted to Dr. H. A. Yamani for very enlightening and fruitful discussions. Many thanks go to Dr. M. S. Abdelmonem for the invaluable help in literature search.

## APPENDIX A: OVERVIEW OF THE RELATIVISTIC J-MATRIX

The J-matrix [26] is an algebraic method of quantum scattering whose structure in function space parallels that of the R-matrix method in configuration space. The perturbing short range potential, $\tilde{V}(r)$, in the R-matrix method is confined to an "$R$-box" in configuration space [i.e. $\tilde{V}(r) = 0$ for $r \geq R$]. While in the J-matrix method it is confined to an "$N$-box" in function space. That is the matrix representation $\tilde{V}_{nm} = 0$ for $n,m \geq N$. In the two methods, the unperturbed (reference) problem is solved analytically enabling scattering calculation over a continuous range of energy despite the fact that confinement in both methods produce discrete energy spectra. The basis $\{\psi_n\}_{n=0}^{\infty}$ of the function space, in the J-matrix method, is chosen such that the matrix representation of the wave operator, $\Im = H_0 - \varepsilon$ , is tridiagonal. $H_0$ is the reference Hamiltonian and $\varepsilon$ is the energy. This has no parallel in the R-matrix method. It restricts the type of $L^2$ bases and limits reference Hamiltonians to those with this type of symmetry that admits such tridiagonal representations. Therefore, the reference wave equation:

$$\Im|\Phi\rangle = \Im\left(\sum_n d_n |\psi_n\rangle\right) = 0 \qquad \text{(A.1)}$$

gives a symmetric three-term recursion relation for the set of expansion coefficients of the spinor wave function, $\{d_n(\varepsilon)\}_{n=0}^{\infty}$. That is

$$\Im_{n,n-1}d_{n-1}+\Im_{n,n}d_n+\Im_{n,n+1}d_{n+1}=0\ ;\quad n\ge 1 \tag{A.2}$$

where $\Im_{nm}$ is the tridiagonal matrix representation of the wave operator, $\left(H_0\right)_{nm}-\varepsilon\,\Omega_{nm}$, and $\Omega$ is the identity operator (i.e. the basis overlap matrix whose elements are $\Omega_{nm}=\langle\psi_n|\psi_m\rangle$ which is also tridiagonal). The solution of the recursion relation (A.2), subject to proper initial conditions, gives two "regularized" solutions of the relativistic reference wave equation (A.1). That is we obtain two sets of expansion coefficients $\{d_n(\varepsilon)\}$ = $\{s_n(\varepsilon)\}$ or $\{c_n(\varepsilon)\}$ giving two wavefunction solutions that are finite at the origin and behave asymptotically as sin($kr$) or cos($kr$), respectively, where $k$ is the energy-dependent wave number. Hence, the name sine-like and cosine-like coefficients for $s_n$ and $c_n$, respectively. The relativistic extension of the method has been developed for two cases of reference Hamiltonians: the free Dirac Hamiltonian [24,27] and the Dirac-Coulomb Hamiltonian [28]. The proper initial relations that complement (A.2) for the two sets of expansion coefficients are [24,28]:

$$\begin{aligned}&\Im_{00}s_0+\Im_{01}s_1=0\\&\Im_{00}c_0+\Im_{01}c_1=-\alpha^2 w/2s_0\end{aligned} \tag{A.3}$$

where $w(\varepsilon)$ is the Wronskian of the regular and irregular solutions of the free Dirac problem. Therefore, for a given $H_0$, basis $\{\psi_n\}_{n=0}^{\infty}$ and initial coefficients $s_0$ and $c_0$, the whole set $\{s_n,c_n\}_{n=0}^{\infty}$ will be determined recursively using (A.3) and (A.2). The recursion relation (A.2) is frequently written in terms of the "J-matrix kinematical coefficients"

$$T_n\equiv\frac{c_n-is_n}{c_n+is_n}\quad;\quad R_{n+1}^{\pm}\equiv\frac{c_{n+1}\pm is_{n+1}}{c_n\pm is_n}\quad;n\ge 0 \tag{A.4}$$

as follows:

$$R_{n+1}^{\pm}=-\frac{1}{\Im_{n,n+1}}\left(\Im_{n,n}+\frac{\Im_{n,n-1}}{R_n^{\pm}}\right)\quad\text{and}\quad T_n=T_{n-1}\frac{R_n^{-}}{R_n^{+}}\quad;\quad n\ge 1 \tag{A.5}$$

It is solved for given initial coefficients $T_0$ and $R_1^{\pm}$.

For the free Dirac Hamiltonian:

$$H_0=\begin{pmatrix}1 & \alpha\left(\frac{\kappa}{r}-\frac{d}{dr}\right)\\ \alpha\left(\frac{\kappa}{r}+\frac{d}{dr}\right) & -1\end{pmatrix} \tag{A.6}$$

the two-component $L^2$ spinor basis, which is relevant to our present treatment, has already been obtained [24,28]. This basis is either of the Laguerre- or oscillator-type. The upper and lower radial components of the Laguerre-type spinor basis are written in terms of the generalized Laguerre polynomials, respectively, as

$$\begin{aligned}&\phi_n(r)=a_n(\lambda r)^{\kappa+1}e^{-\lambda r/2}L_n^{2\kappa+1}(\lambda r)\\&\theta_n(r)=\frac{\lambda C}{2}a_n(\lambda r)^{\kappa}e^{-\lambda r/2}\left[(2\kappa+n+1)L_n^{2\kappa}(\lambda r)+(n+1)L_{n+1}^{2\kappa}(\lambda r)\right]\end{aligned} \tag{A.7}$$

where $\lambda$ is a scale parameter, $C$ is the small component strength parameter, $L_n^{\nu}(x)$ is the generalized Laguerre polynomial, and the normalization constant is

$$a_n = \sqrt{\frac{\lambda\Gamma(n+1)}{\Gamma(2\kappa+n+2)}} \tag{A.8}$$

The matrix representation of the reference Hamiltonian $H_0$ in this basis is tridiagonal and has the following elements:

$$\begin{aligned}
\left(H_0\right)_{n,n} &= 2(\kappa+n+1)\left[1-\left(\lambda C/2\right)^2\left(1-2\alpha/C\right)\right] \\
\left(H_0\right)_{n,n+1} &= -\sqrt{(n+1)(2\kappa+n+2)}\left[1+\left(\lambda C/2\right)^2\left(1-2\alpha/C\right)\right] \\
\left(H_0\right)_{n,n-1} &= -\sqrt{n(2\kappa+n+1)}\left[1+\left(\lambda C/2\right)^2\left(1-2\alpha/C\right)\right]
\end{aligned} \tag{A.9}$$

The identity operator $\Omega$ is also tridiagonal:

$$\begin{aligned}
\Omega_{n,n} &= 2(\kappa+n+1)\left[1+\left(\lambda C/2\right)^2\right] \\
\Omega_{n,n+1} &= -\sqrt{(n+1)(2\kappa+n+2)}\left[1-\left(\lambda C/2\right)^2\right] \\
\Omega_{n,n-1} &= -\sqrt{n(2\kappa+n+1)}\left[1-\left(\lambda C/2\right)^2\right]
\end{aligned} \tag{A.10}$$

The initial J-matrix kinematical coefficients in this basis for the case where $\kappa = 0$, which is relevant to our work here, are [24,26]:

$$T_0 = e^{2i\omega} \quad ; \quad R_1^{\pm} = \frac{1}{\sqrt{2}} e^{\mp i\omega} \tag{A.11}$$

where

$$\cos(\omega) = \frac{\left[k(\varepsilon)/\lambda\right]^2 - 1/4}{\left[k(\varepsilon)/\lambda\right]^2 + 1/4} \tag{A.12}$$

and

$$k(\varepsilon) = \sqrt{\frac{-1}{C^2}\frac{\varepsilon-1}{\varepsilon-1+2(1-\alpha/C)}} \tag{A.13}$$

The nonrelativistic limit is obtained [24,28] by taking

$$\begin{aligned}
&\alpha \to 0 \\
&C = \alpha/2
\end{aligned} \tag{A.14}$$

In the oscillator-type spinor basis, however, the upper and lower radial components are written, respectively, as

$$\begin{aligned}
\phi_n(r) &= a_n(\lambda r)^{\kappa+1} e^{-\lambda^2 r^2/2} L_n^{\kappa+1/2}(\lambda^2 r^2) \\
\theta_n(r) &= \lambda C a_n(\lambda r)^{\kappa} e^{-\lambda^2 r^2/2}\left[(n+\kappa+1/2)L_n^{\kappa-1/2}(\lambda^2 r^2)+(n+1)L_{n+1}^{\kappa-1/2}(\lambda^2 r^2)\right]
\end{aligned} \tag{A.15}$$

where the normalization constant, on the other hand, is

$$a_n = \sqrt{\frac{2\lambda\Gamma(n+1)}{\Gamma(n+\kappa+3/2)}} \tag{A.16}$$

The matrix representation of the reference Hamiltonian $H_0$ in this basis is also tridiagonal with the following elements:

$$\begin{aligned}
\left(H_0\right)_{nn} &= 1+\lambda^2 C^2(-1+2\alpha/C)(2n+\kappa+3/2) \\
\left(H_0\right)_{n,n-1} &= \lambda^2 C^2(-1+2\alpha/C)\sqrt{n(n+\kappa+1/2)} \\
\left(H_0\right)_{n,n+1} &= \lambda^2 C^2(-1+2\alpha/C)\sqrt{(n+1)(n+\kappa+3/2)}
\end{aligned} \tag{A.17}$$

The tridiagonal overlap matrix has the following elements:

$$\Omega_{nn} = 1 + \lambda^2 C^2 (2n + \kappa + 3/2)$$
$$\Omega_{n,n-1} = \lambda^2 C^2 \sqrt{n(n + \kappa + 1/2)} \qquad \text{(A.18)}$$
$$\Omega_{n,n+1} = \lambda^2 C^2 \sqrt{(n+1)(n + \kappa + 3/2)}$$

The initial sine-like and cosine-like coefficients in this basis for the case $\kappa = 0$ are [24,26]:

$$\left.\begin{matrix} s_0(\varepsilon) \\ s_1(\varepsilon) \end{matrix}\right\} = \pi^{1/4} \sqrt{\frac{2}{\lambda}\frac{k}{\lambda}} e^{-k^2/2\lambda^2} \begin{cases} 1 \\ \dfrac{k^2/\lambda^2 - 3/2}{\sqrt{3}} \end{cases} \qquad \text{(A.19.1)}$$

$$\left.\begin{matrix} c_0(\varepsilon) \\ c_1(\varepsilon) \end{matrix}\right\} = \pi^{-1/4} \sqrt{\frac{2}{\lambda}} e^{-k^2/2\lambda^2} \begin{cases} {}_1F_1\left(-1/2;1/2;k^2/\lambda^2\right) \\ -\dfrac{{}_1F_1\left(-3/2;1/2;k^2/\lambda^2\right)}{\sqrt{3}} \end{cases} \qquad \text{(A.19.2)}$$

where ${}_1F_1(a;b;z)$ is the confluent hypergeometric function and $k(\varepsilon)$ is as defined in (A.13). The nonrelativistic limit is also obtained by using (A.14).

## APPENDIX B: CALCULATING THE PHASE SHIFT FOR THE CASE $\nu = 1$

The transformation (2.18) takes the initial conditions in (2.17) to

$$\begin{pmatrix} \Im_{00} & \Im_{01} & 0 \\ \Im_{10} & \Im_{11} & \Im_{12} \end{pmatrix} \begin{pmatrix} \eta e^{i\xi} g_0^+ \\ e^{i\tau} g_1^+ \\ e^{i\tau} g_2^+ \end{pmatrix} = -\begin{pmatrix} V_{00} & V_{01} \\ V_{10} & V_{11} \end{pmatrix} \begin{pmatrix} \eta e^{i\xi} g_0^+ \\ e^{i\tau} g_1^+ \end{pmatrix} + \frac{-i\alpha^2 w / \eta e^{i\xi} g_0^+}{1 - e^{-2i\xi} T_0} \begin{pmatrix} 1 \\ 0 \end{pmatrix} \qquad \text{(B.1)}$$

Using (2.12), we can write

$$\begin{aligned} \Im_{00} \eta e^{i\xi} g_0^+ + \Im_{01} e^{i\tau} g_1^+ &= (\eta e^{i\xi} - e^{i\tau}) \Im_{00} g_0^+ + e^{i\tau} (\Im_{00} g_0^+ + \Im_{01} g_1^+) \\ &= (\eta e^{i\xi} - e^{i\tau}) \Im_{00} g_0^+ + \frac{-i\alpha^2 w e^{i\tau} / g_0^+}{1 - T_0} \end{aligned} \qquad \text{(B.2)}$$

We can also write

$$\begin{aligned} \Im_{10} \eta e^{i\xi} g_0^+ + e^{i\tau} (\Im_{11} g_1^+ + \Im_{12} g_2^+) &= (\eta e^{i\xi} - e^{i\tau}) \Im_{10} g_0^+ + e^{i\tau} (\Im_{10} g_0^+ + \Im_{11} g_1^+ + \Im_{12} g_2^+) \\ &= (\eta e^{i\xi} - e^{i\tau}) \Im_{10} g_0^+ = (\eta e^{i\xi} - e^{i\tau}) \Im_{01} g_0^+ \end{aligned} \qquad \text{(B.3)}$$

Using these results, we can write equation (B.1) as

$$\begin{aligned} (\eta e^{i\xi} - e^{i\tau}) g_0^+ \begin{pmatrix} \Im_{00} \\ \Im_{01} \end{pmatrix} + \frac{-i\alpha^2 w e^{i\tau} / g_0^+}{1 - T_0} \begin{pmatrix} 1 \\ 0 \end{pmatrix} &= \\ -\begin{pmatrix} V_{00} & V_{01} \\ V_{10} & V_{11} \end{pmatrix} \begin{pmatrix} \eta e^{i\xi} g_0^+ \\ e^{i\tau} g_1^+ \end{pmatrix} &+ \frac{-i\alpha^2 w / \eta e^{i\xi} g_0^+}{1 - e^{-2i\xi} T_0} \begin{pmatrix} 1 \\ 0 \end{pmatrix} \end{aligned} \qquad \text{(B.4)}$$

The second raw yields the following equation

$$\eta e^{i\xi} g_0^+ (\Im_{01} + V_{01}) = e^{i\tau} (\Im_{01} g_0^+ - V_{11} g_1^+) = e^{i\tau} g_0^+ (\Im_{01} - V_{11} \hat{R}_1^+) \qquad \text{(B.5)}$$

giving

$$\eta e^{i\xi} = e^{i\tau} \frac{\Im_{01} - V_{11} R_1^+}{\Im_{01} + V_{01}} \qquad \text{(B.6)}$$

Therefore, we conclude that

$$\eta = \left| \frac{\Im_{01} - V_{11} R_1^+}{\Im_{01} + V_{01}} \right| \tag{B.7}$$

and

$$e^{i\xi} = e^{i\tau} e^{i\zeta} \tag{B.8}$$

where

$$\zeta = \arg\left( \frac{\Im_{01} - V_{11} R_1^+}{\Im_{01} + V_{01}} \right) \tag{B.9}$$

The first raw of (B.4) gives the following equation

$$\Im_{00}(\eta e^{i\zeta} - 1) - \frac{i\alpha^2 w}{(g_0^+)^2} \frac{1}{1 - T_0} = -\left( V_{00} \eta e^{i\zeta} + V_{01} R_1^+ \right) - \frac{i\alpha^2 w}{\eta e^{i\zeta} (g_0^+)^2} \frac{1}{e^{2i\tau} - e^{-2i\zeta} T_0} \tag{B.10}$$

Equation (2.12) also gives

$$(g_0^+)^2 = \frac{-i\alpha^2 w}{1 - T_0} \left( \Im_{00} + \Im_{01} R_1^+ \right)^{-1} \tag{B.11}$$

Substituting in (B.10), we obtain

$$\frac{\Im_{00} + \Im_{01} R_1^+}{e^{2i\tau} - T_0 e^{-2i\zeta}} = \frac{\eta e^{i\zeta}}{1 - T_0} \left[ \eta e^{i\zeta} \left( \Im_{00} + V_{00} \right) + R_1^+ \left( \Im_{01} + V_{01} \right) \right] \tag{B.12}$$

giving us, finally, the sought after scattering matrix:

$$\begin{aligned} e^{2i\tau} = {} & T_0 e^{-2i\zeta} + \left(1 - T_0\right) \left( \Im_{01} + V_{01} \right) \left( \frac{\Im_{00} + \Im_{01} R_1^+}{\Im_{01} - V_{11} R_1^+} \right) \times \\ & \left[ \left( \Im_{01} - V_{11} R_1^+ \right) \frac{\Im_{00} + V_{00}}{\Im_{01} + V_{01}} + R_1^+ \left( \Im_{01} + V_{01} \right) \right]^{-1} \end{aligned} \tag{B.13}$$

**FIGURES CAPTION:**

**Figure (1):** The relativistic (nonrelativistic) phase shift, in radians, as solid (dotted) line vs. the nonrelativistic energy, in atomic units. The separable potential is of the type $r^{\nu-1}e^{-\lambda r/2}$ with the coupling parameters in (4.1). The rest of the parameter values are taken as:

(a) $\nu = 0$, $\lambda = 1.0$, $\alpha = 0.4$, $C = \alpha / 3$
(b) $\nu = 1$, $\lambda = 1.5$, $\alpha = 0.5$, $C = \alpha / 3$
(c) $\nu = 2$, $\lambda = 2.0$, $\alpha = 0.5$, $C = \alpha / 3$
(d) $\nu = 3$, $\lambda = 2.5$, $\alpha = 0.2$, $C = \alpha / 3$

**Figure (2):** The relativistic (nonrelativistic) phase shift, in radians, as solid (dotted) line vs. the nonrelativistic energy, in atomic units. The separable potential is of the type $r^{2\nu}e^{-\lambda^2 r^2/2}$ with the coupling parameters in (4.1). The rest of the parameter values are taken as:

(a) $\nu = 0$, $\lambda = 1.0$, $\alpha = 0.5$, $C = \alpha / 3$
(b) $\nu = 1$, $\lambda = 1.1$, $\alpha = 0.4$, $C = \alpha / 3$
(c) $\nu = 2$, $\lambda = 1.4$, $\alpha = 0.3$, $C = \alpha / 4$
(d) $\nu = 3$, $\lambda = 1.5$, $\alpha = 0.3$, $C = \alpha / 4$

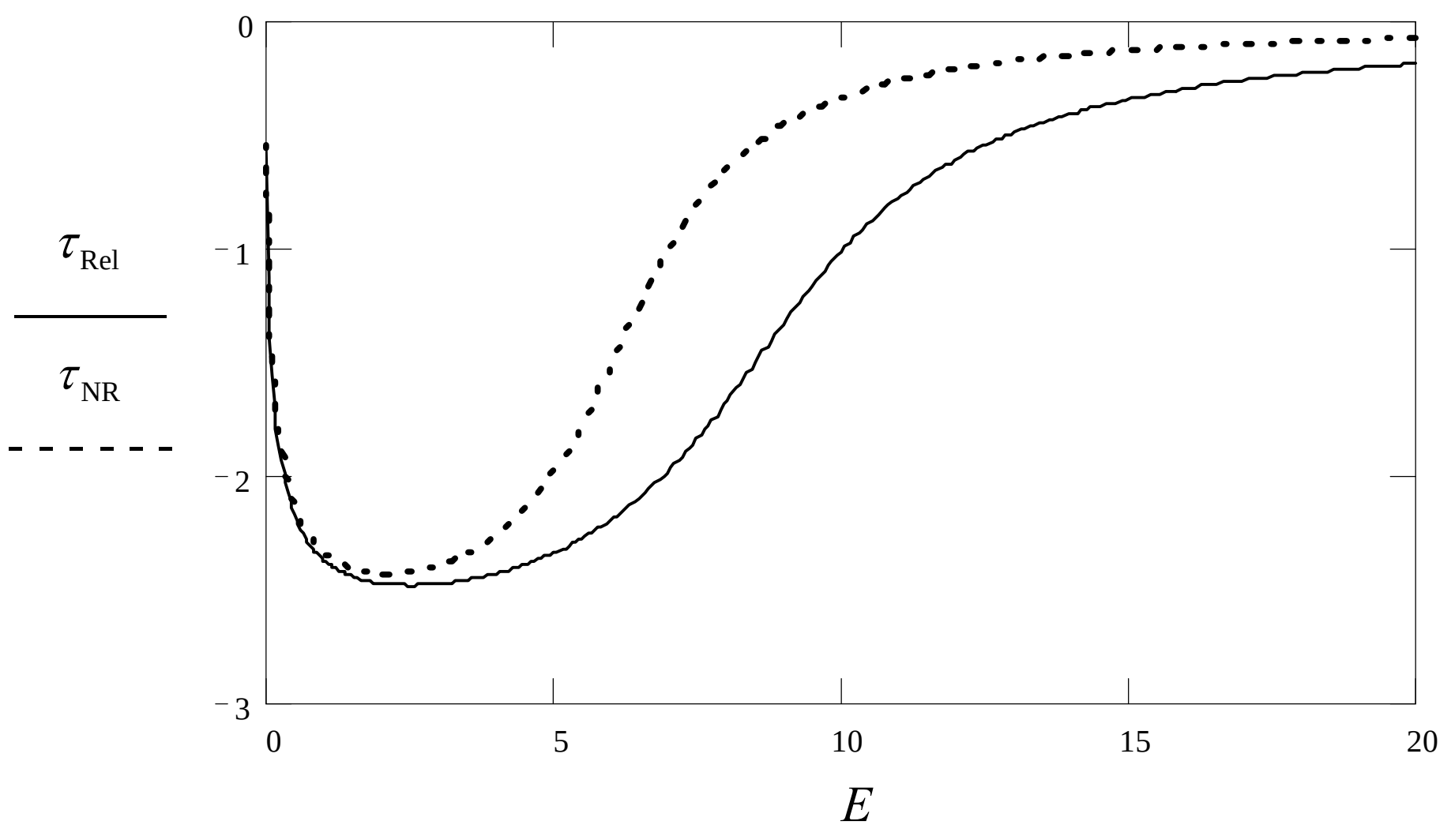

Figure (1-a)

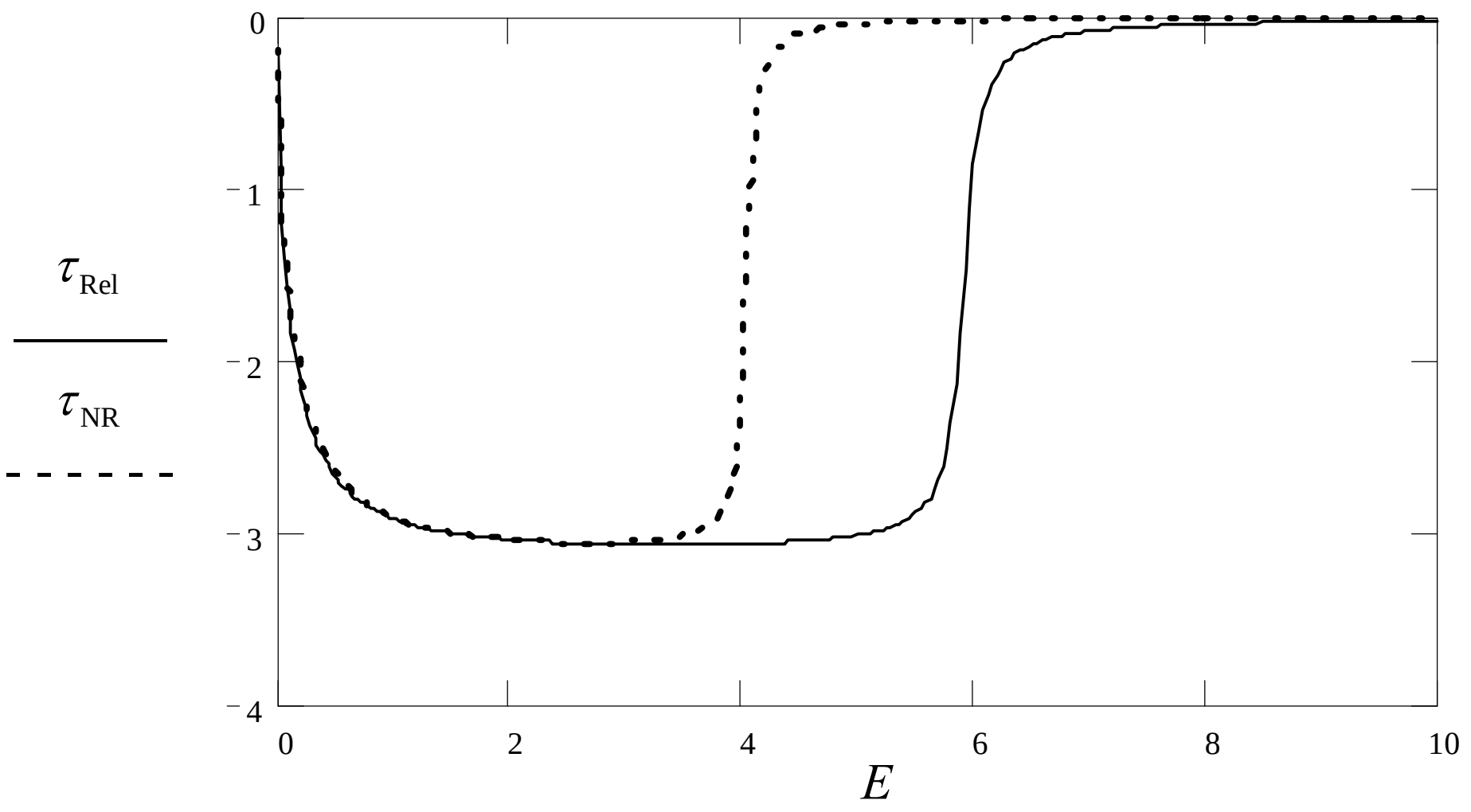

Figure (1-b)

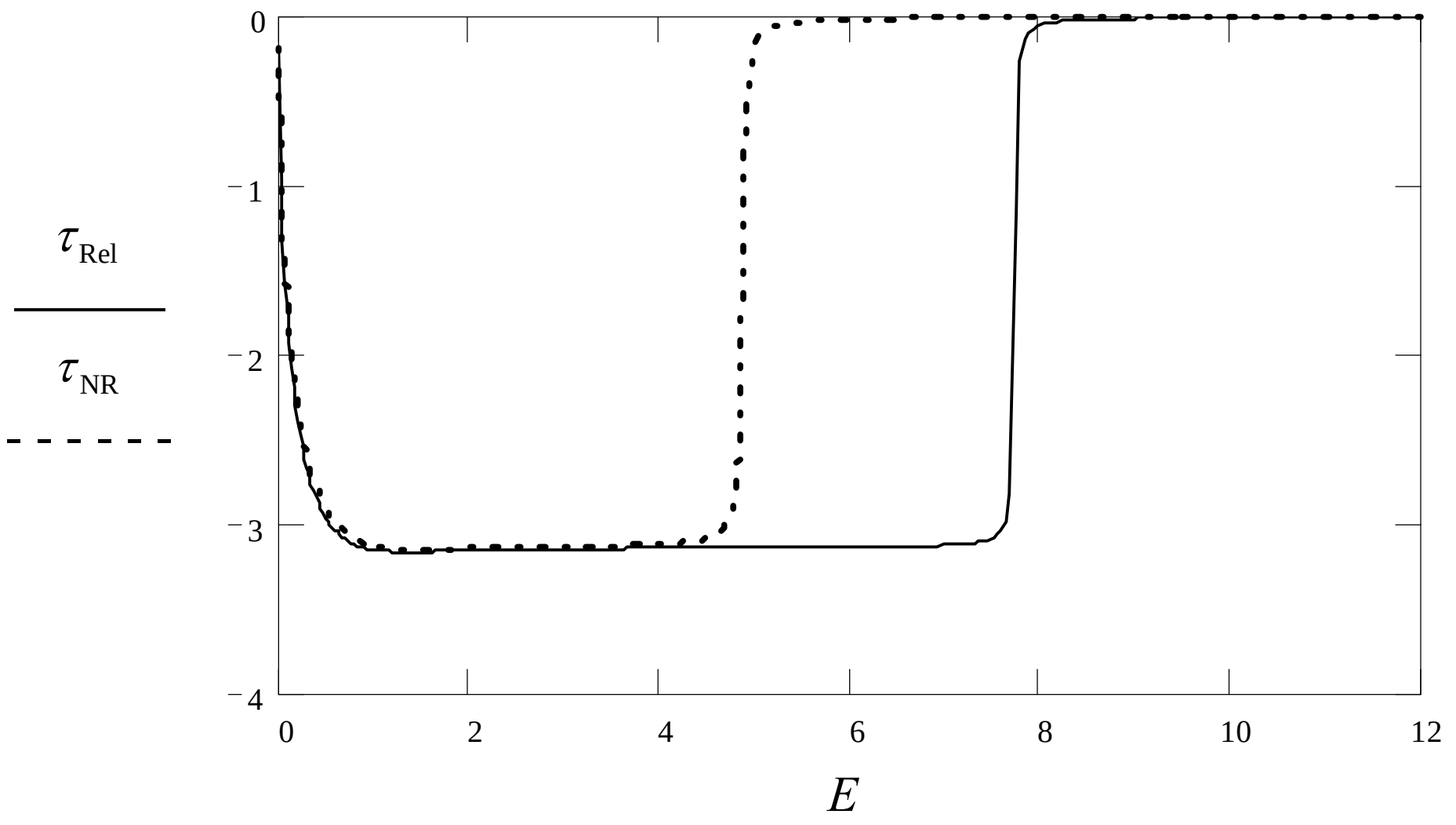

Figure (1-c)

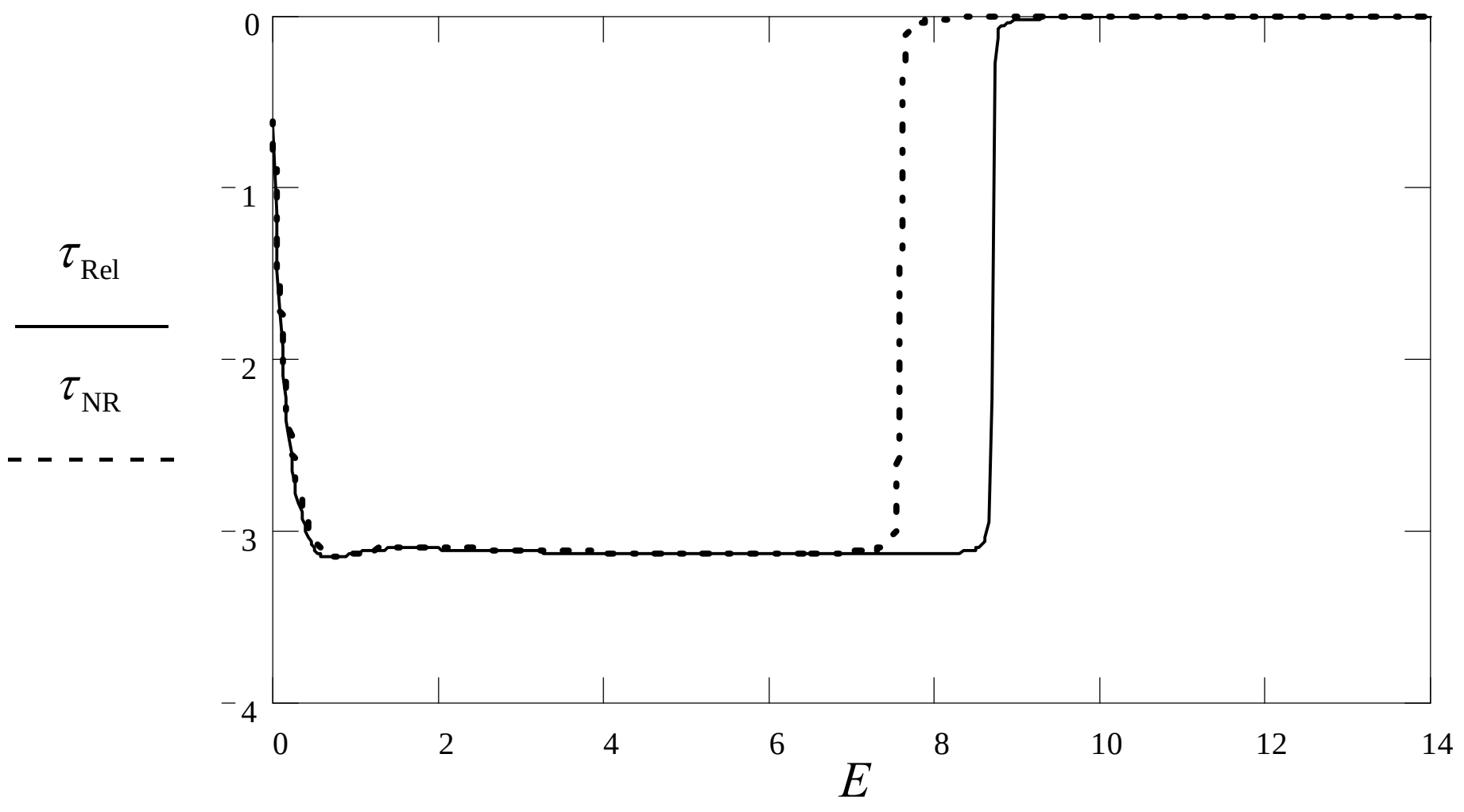

Figure (1-d)

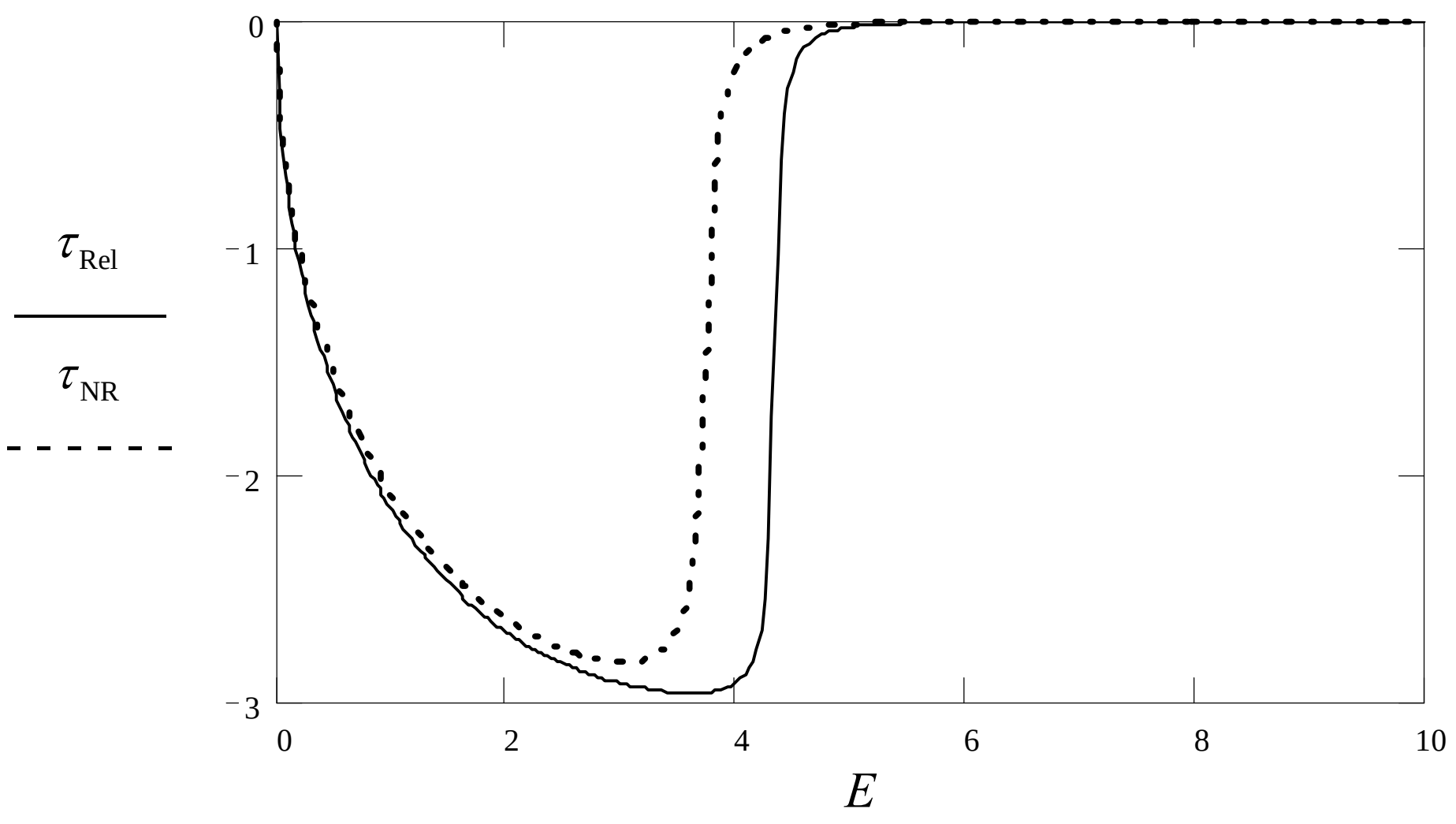

Figure (2-a)

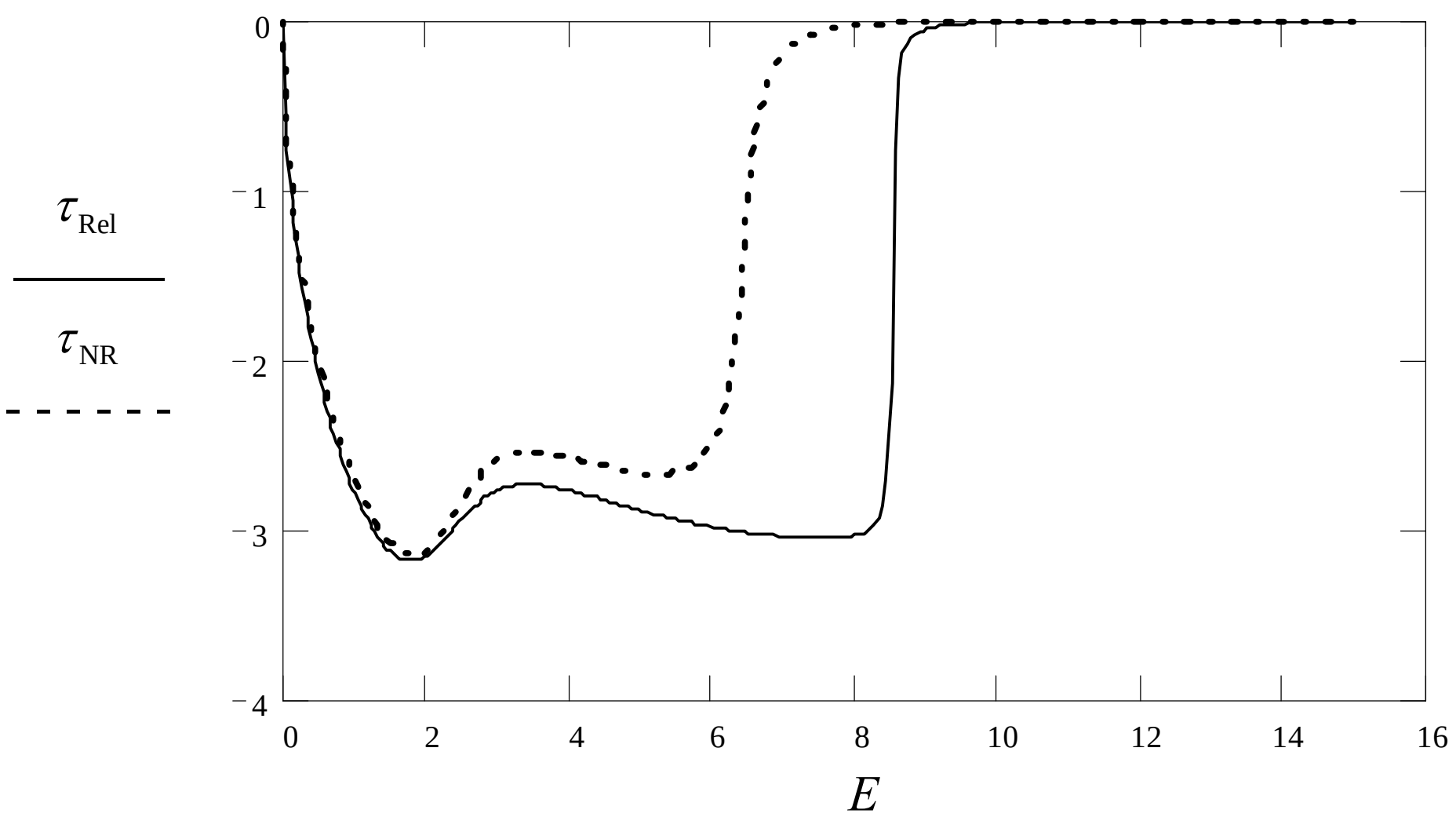

Figure (2-b)

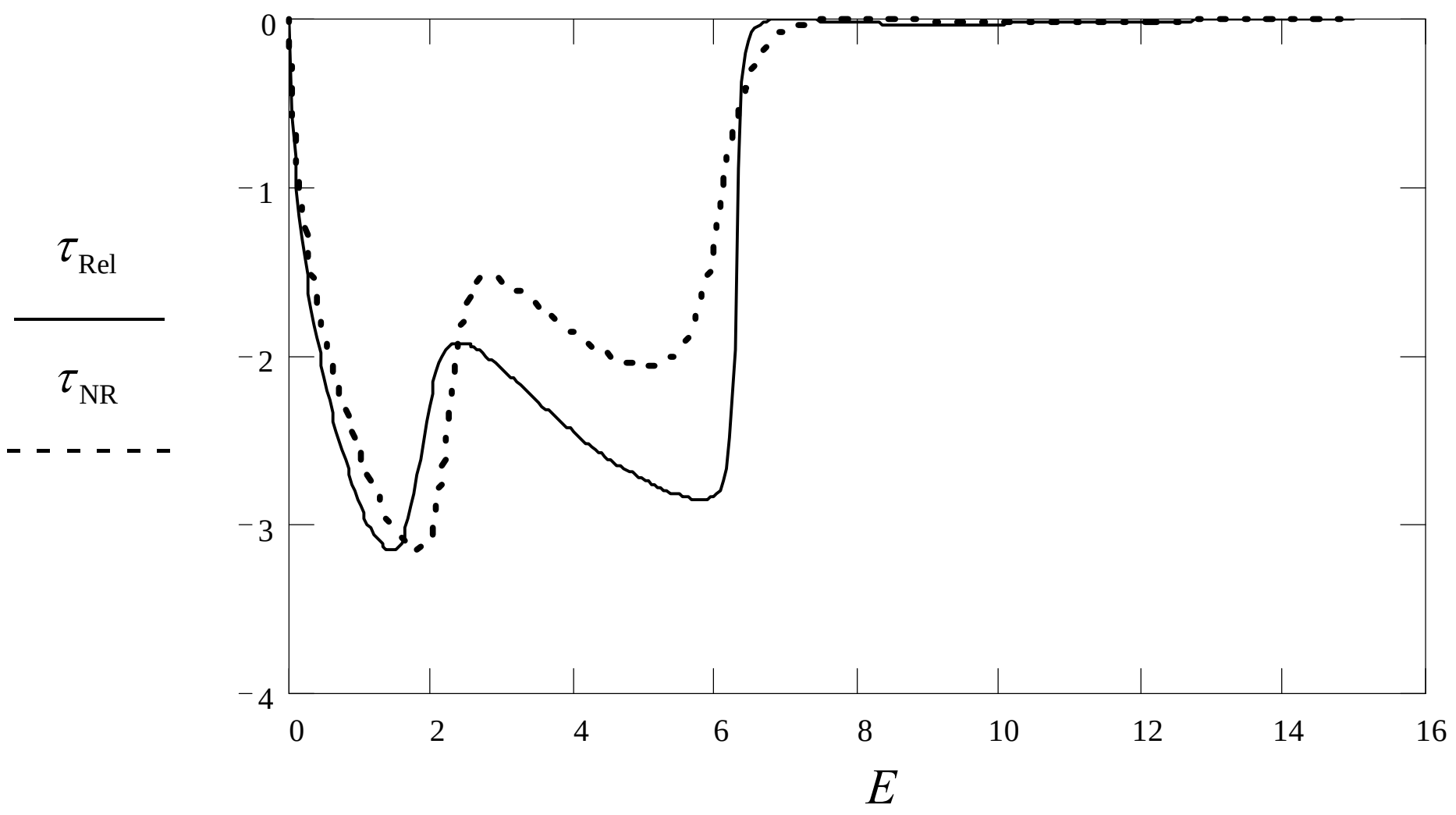

Figure (2-c)

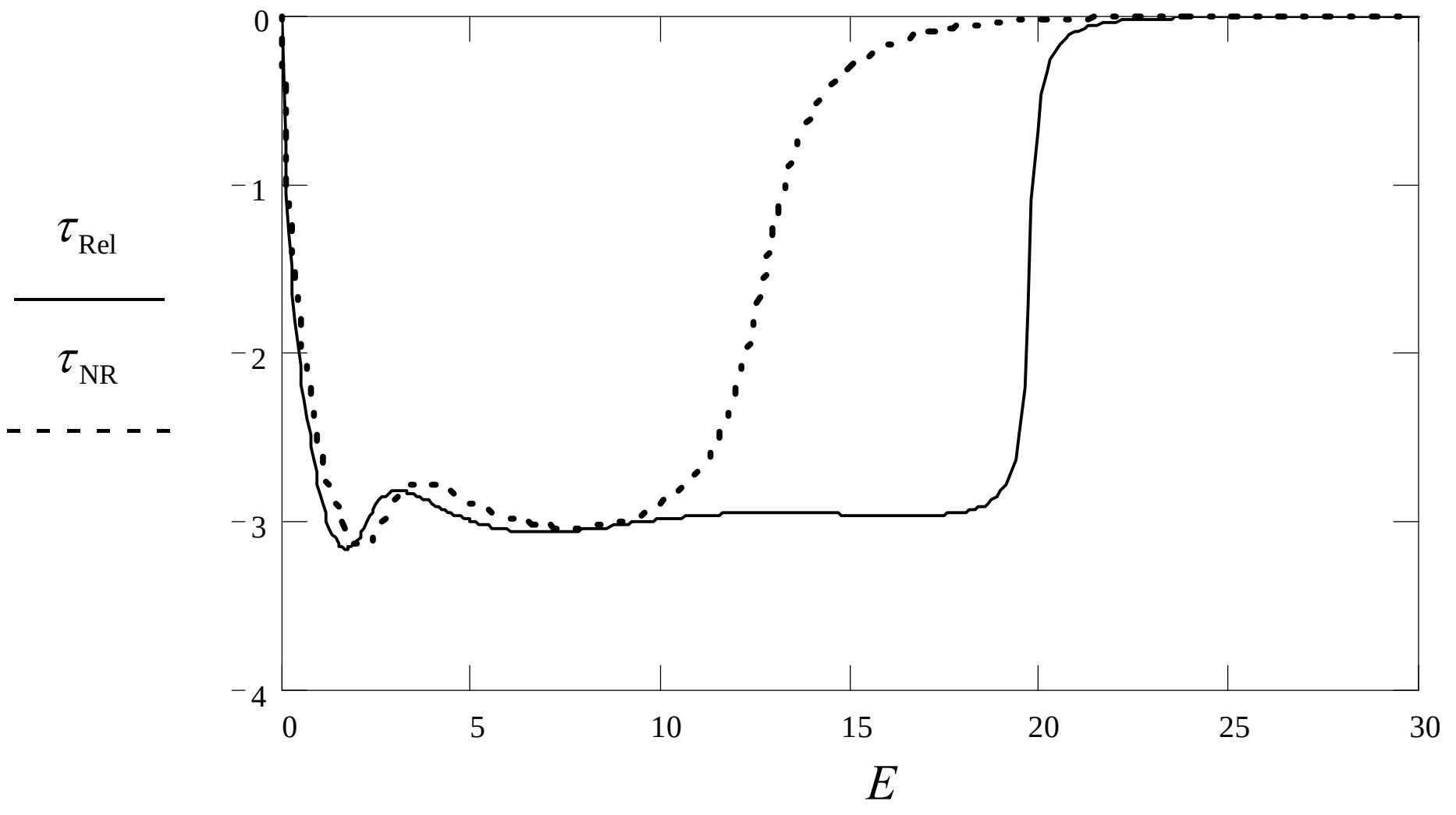

Figure (2-d)